\documentclass[10pt,conference]{IEEEtran}
\usepackage{epsf,cite,amsmath,amscd,amssymb,graphicx,latexsym,multicol}

\begin{document}

\bibliographystyle{ieeetr}
\title{Energy Efficiency and Delay Quality-of-Service\\ in Wireless Networks}

\author{\authorblockN{Farhad Meshkati, H. Vincent Poor, Stuart C.
Schwartz}
\authorblockA{Department of Electrical Engineering\\Princeton University\\
Princeton, NJ 08544 USA\\
Email: \{meshkati, poor, stuart\}@princeton.edu} \and
\authorblockN{Radu V. Balan}
\authorblockA{Siemens Corporate Research\\755 College Road East\\Princeton, NJ 08540
USA\\
Email: radu.balan@siemens.com}
}

\newtheorem{proposition}{Proposition}
\newenvironment{thmproof}[1]
{\noindent\hspace{2em}{\it #1 }}
{\hspace*{\fill}~\QED\par\endtrivlist\unskip}

\centerfigcaptionstrue

\maketitle

\begin{abstract}
The energy-delay tradeoffs in wireless networks are studied using a
game-theoretic framework. A multi-class multiple-access network is
considered in which users choose their transmit powers, and possibly
transmission rates, in a distributed manner to maximize their own
utilities while satisfying their delay quality-of-service (QoS)
requirements. The utility function considered here measures the
number of reliable bits transmitted per Joule of energy consumed and
is particularly useful for energy-constrained networks. The Nash
equilibrium solution for the proposed non-cooperative game is
presented and closed-form expressions for the users' utilities at
equilibrium are obtained. Based on this, the losses in energy
efficiency and network capacity due to presence of delay-sensitive
users are quantified. The analysis is extended to the scenario where
the QoS requirements include both the average source rate and a
bound on the average total delay (including queuing delay). It is
shown that the incoming traffic rate and the delay constraint of a
user translate into a ``size" for the user, which is an indication
of the amount of resources consumed by the user. Using this
framework, the tradeoffs among throughput, delay, network capacity
and energy efficiency are also quantified.
\end{abstract}

\section{Introduction}
Future wireless networks are expected to support a variety of
services with diverse quality of service (QoS) requirements. For
example, a mixture of delay-sensitive and delay-tolerant users could
exist in the same network. At the same time, most of the user
terminals in a wireless network are battery-powered. As a result,
energy efficiency is also crucial in design of wireless networks.
Therefore, the objective is to use the radio resources (e.g., power
and bandwidth) as efficiently as possible and at the same time
satisfy the QoS requirements of the users in the network.

In this work, we study the tradeoffs between energy efficiency and
delay QoS using a game-theoretic framework. We consider a
multiple-access network in which each user seeks to locally choose
its transmit power, and possibly its transmission rate, in order to
maximize its own utility (in bits per Joule) and at the same time
satisfy its delay QoS requirements. The strategy chosen by each user
affects the other users through multiple-access interference. The
study of the tradeoffs between energy efficiency and delay has
recently attracted considerable attention (see for example
\cite{Collins99, Prabhakar01, Berry02, Uysal02, Fu03, Coleman04}).
Our non-cooperative game-theoretic approach allows us to study the
energy efficiency-delay tradeoffs in a multiuser competitive
setting. Using this framework, we quantify the loss in energy
efficiency due to the presence of delay-sensitive users in the
network and analyze the tradeoffs among throughput, delay, network
capacity and energy efficiency.

The remainder of the paper is organized as follows. In
Section~\ref{system model}, we present the system model and define
the user utility function. The delay model for the infinite backlog
case is given in Section~\ref{delay model infinite}. The proposed
delay-constrained power control game and its Nash equilibrium
solution are presented in Section~\ref{PCG}.  In
Section~\ref{multiclass}, we give explicit expressions for the
utilities achieved at Nash equilibrium for a multi-class network.
The delay model for the finite backlog case is given in
Section~\ref{delay model finite}. In Section~\ref{PRCG}, we propose
a delay-constrained power and rate control game and give its Nash
equilibrium solution. Numerical results and conclusions are given in
Sections~\ref{numerical results} and~\ref{conclusion}, respectively.

\section{System Model}\label{system model}
We consider a synchronous direct-sequence
code-division-multiple-access (DS-CDMA) network with $K$ users and
processing gain $N$ (defined as the ratio of symbol duration to chip
duration). We assume that all $K$ user terminals transmit to a
receiver at a common concentration point. The received signal at the
access point sampled at the chip rate over one symbol duration can
be expressed as
\begin{equation}\label{eq1}
   {\mathbf{r}} = \sum_{k=1}^{K} \sqrt{p_k} h_k \ b_k {\mathbf{s}}_k +
   {\mathbf{w}} ,
\end{equation}
where $p_k$, $h_k$, $b_k$ and ${\mathbf{s}}_k$ are the transmit
power, channel gain, transmitted bit and spreading sequence of the
$k^{th}$ user, respectively, and $\mathbf{w}$ is the noise vector
which is assumed to be Gaussian with mean $\mathbf{0}$ and
covariance $\sigma^2 \mathbf{I}$. We assume random spreading
sequences for all users.

We assume that data arrives at the user terminal in the form of
$M$-bit packets. The user transmits the arriving packets at a rate
$R_k$ (bps) and with a transmit power equal to $p_k$ Watts. We
consider an automatic-repeat-request (ARQ) mechanism in which the
user keeps retransmitting a packet until the packet is received at
the access point without any errors. Let us define the utility
function of a user to be the ratio of its goodput to its transmit
power, i.e.,
\begin{equation}\label{eq2}
   u_k = \frac{T_k}{p_k} \ .
\end{equation}
Goodput is the net number of information bits that are transmitted
without error per unit time and is expressed as
\begin{equation}\label{eq3}
   T_k = R_k f(\gamma_k)
\end{equation}
where $\gamma_k$ is the output SIR for user $k$ and $f(\gamma_k)$ is
the ``efficiency function" which represents the packet success rate
(PSR). We assume $f(\gamma)$ to be continuous, increasing and
S-shaped\footnote{An increasing function is S-shaped if there is a
point above which the function is concave, and below which the
function is convex.} (sigmoidal) with $f(\infty)=1$. This is a valid
assumption for many practical scenarios as long as the packet size
is reasonably large (e.g., $M=100$ bits). We also require that
$f(0)=0$ to ensure that $u_k=0$ when $p_k=0$. In general, the
efficiency function depends on the modulation, coding and packet
size. A more detailed discussion of the efficiency function can be
found in \cite{Meshkati_TCOMM}. Based on (\ref{eq2}) and
(\ref{eq3}), the utility function for user $k$ can be written as
\begin{equation}\label{eq4}
   u_k = R_k \frac{f(\gamma_k)}{p_k}\ .
\end{equation}
This utility function, which has units of \emph{bits/Joule},
captures very well the tradeoff between throughput and battery life,
and is particularly suitable for energy-constrained networks.

\section{Delay Model for the Infinite Backlog Case}\label{delay model infinite}

Let us for now focus on the case in which there are infinitely many
packets to be transmitted by each user. For this case, we
concentrate on the transmission delay. Let $X$ represent the
(random) number of transmissions required for a packet to be
received without any errors. The assumption is that if a packet has
one or more errors, it will be retransmitted. We also assume that
retransmissions are independent from each other. It is clear that
the transmission delay for a packet is directly proportional to $X$.
Since the packet success rate is given by the efficiency function
$f(\gamma)$, the probability that exactly $m$ transmissions are
required for the successful transmission of a packet is given by
\begin{equation}\label{eq5}
\textrm{Pr}\{X=m\}= f(\gamma) \left( 1-f(\gamma) \right)^{m-1} \ .
\end{equation}
We model the delay requirements of a particular as a pair
$(L,\beta)$, where
\begin{equation}\label{eq6}
\textrm{Pr}\{X\leq L\}\geq \beta .
\end{equation}
In other words, we would like the number of transmissions to be at
most $L$ with a probability larger than or equal to $\beta$. Note
that \eqref{eq6} can equivalently be represented as an upper bound
on the delay outage probability. Based on \eqref{eq5}, it can be
shown that the delay constraint in \eqref{eq6} is equivalent to
\begin{equation}\label{eq7}
   f(\gamma) \geq \tilde{\eta}(L,\beta) ,
\end{equation}
where
\begin{equation}\label{eq7b}
   \tilde{\eta}(L,\beta)=1-(1-\beta)^{\frac{1}{L}} .
\end{equation}
Since $f(\gamma)$ is an increasing function of $\gamma$, we can
equivalently express \eqref{eq7} as
\begin{equation}\label{eq8}
   \gamma \geq \tilde{\gamma}
\end{equation}
where $\tilde{\gamma}= f^{-1} \left( \tilde{\eta} (L,\beta)\right)$.
Therefore, the delay constraint in \eqref{eq6} translates into a
lower bound on the output SIR. Since different users could have
different delay requirements, $\tilde{\gamma}$ is user dependent. We
make this explicit by writing
\begin{equation}\label{eq10}
   \tilde{\gamma}_k= f^{-1} \left( \tilde{\eta}_k \right)
\end{equation}
where $\tilde{\eta}_k=1-(1-\beta_k)^{\frac{1}{L_k}}$. A more
stringent delay requirement, i.e., a smaller $L$ and/or a larger
$\beta$, will result in a higher value for $\tilde{\gamma}$.

\section{Power Control Game with Delay Constraints}\label{PCG}

Consider the non-cooperative power control game (PCG)
$\tilde{G}=[{\mathcal{K}}, \{\tilde{A}_k \}, \{{u}_k \}]$ where
${\mathcal{K}}=\{1, ... , K \}$, and $\tilde{A}_k=[0,P_{max}]$,
which is the strategy set for the $k^{th}$ user and $u_k$ is the
utility function given by \eqref{eq4}. Here, $P_{max}$ is the
maximum allowed power for transmission. Each user chooses its
transmit power in order to maximize its own utility and at the same
time satisfy its delay requirements. We have shown in Section
\ref{delay model infinite} that the delay requirements of a user
translate into a lower bound on the user's output SIR. Hence, the
resulting delay-constrained power control game can be expressed as
\begin{equation}\label{eq13}
   \max_{p_k} \ u_k  \ \ \ \textrm{s.t.} \ \ \ \gamma_k \geq \tilde{\gamma}_k \ \ \textrm{for} \ \
   k=1,...,K.
\end{equation}

We assume that only those users whose delay requirements can be met
are admitted into the network. For example, for the conventional
matched filter, this translates into having $\sum_{k=1}^K
\frac{1}{1+\frac{N}{\tilde{\gamma}_k}} <1$. This assumption makes
sense because admitting a user that cannot meet its delay
requirement only causes unnecessary interference for other users.
The Nash equilibrium for the proposed game is a set strategies
(power levels) for which no user can unilaterally improve its own
(delay-constrained) utility function. We now state the following
proposition.\vspace{0.1cm}

\begin{proposition}\label{prop1}
The Nash equilibrium for the proposed delay-constrained power
control game is given by $\tilde{p}_k^*=\min \{p_k^*, P_{max}\}$,
for $k=1, \cdots, K$, where $p_k^*$ is the transmit power that
results in an output SIR equal to $\tilde{\gamma}_k^*$ with
$\tilde{\gamma}_k^*= \max\{\tilde{\gamma}_k,\gamma^*\}$. Here,
$\tilde{\gamma}_k$ is given by \eqref{eq10} and $\gamma^*$ is the
(positive) solution of $f(\gamma)=\gamma f'(\gamma)$. Furthermore,
this equilibrium is unique.
\end{proposition}
\begin{thmproof}{Proof:} See \cite{MeshkatiISIT} for the proof.\vspace{0.1cm}
\end{thmproof}

The above proposition suggests that at Nash equilibrium, the output
SIR for user $k$ is $\tilde{\gamma}^*_k$, where $\tilde{\gamma}^*_k$
depends on the efficiency function through $\gamma^*$ as well as
user $k$'s delay constraint through $\tilde{\gamma}_k$. Note that
this result does not depend on the choice of the receiver and is
valid for all linear receivers.

\section{Multi-class Networks}\label{multiclass}

Let us now consider a network with $C$ classes of users. The
assumption is that all the users in the same class have the same
delay requirements characterized by the corresponding $L$ and
$\beta$. Based on Proposition \ref{prop1}, at Nash equilibrium, all
the users in class $c$ will have the same output SIR,
$\tilde{\gamma}^{* (c)}= \max\{\tilde{\gamma}^{(c)},\gamma^*\}$,
where $\tilde{\gamma}^{(c)}=f^{-1} \left( \tilde{\eta}^{(c)}
\right)$. The goal is to quantify the effect of delay constraints on
the energy efficiency of the network or equivalently on the users'
utilities.

In order to obtain explicit expressions for the utilities achieved
at equilibrium, we use a large-system analysis. We consider the
asymptotic case in which $K, N \rightarrow \infty $ and $\frac{K}{N}
\rightarrow \alpha < \infty$. This allows us to write SIR
expressions that are independent of the spreading sequences of the
users. Let $K^{(c)}$ be the number of users in class $c$, and define
$\alpha^{(c)}=\lim_{K,N\rightarrow \infty} \frac{K^{(c)}}{N}$.
Therefore, we have $\sum_{c=1}^{C} \alpha^{(c)} = \alpha$. It can be
shown that \cite{MeshkatiISIT} for the matched filter, the
decorrelator, and the linear minimum-mean-square-error (MMSE)
detector, the utilities achieved at the Nash equilibrium are given
by
{\small{\begin{eqnarray} 
    &u_k^{MF}& = \frac{R_k h_k^2}{\sigma^2}\left( 1 - \sum_{c=1}^{C}
    \alpha^{(c)} \tilde{\gamma}^{* (c)}\right) \frac{ f(\tilde{\gamma}^{* (c)})}{ \tilde{\gamma}^{* (c)}} \nonumber\\
    && \hspace{2.5cm} {\textrm{for}} \ \ \sum_{c=1}^{C} \alpha^{(c)} \tilde{\gamma}^{* (c)} < 1, \
    \label{eq21-1}\\ &&\nonumber\\
   &u_k^{DE}& = \frac{R_k h_k^2}{\sigma^2} \left( 1-\sum_{c=1}^{C}
    \alpha^{(c)} \right)  \frac{ f(\tilde{\gamma}^{* (c)})}{ \tilde{\gamma}^{* (c)}} \nonumber\\
    && \hspace{3cm} {\textrm{for}} \ \ \  \sum_{c=1}^{C} \alpha^{(c)}  < 1, \
    \label{eq21-2}\\
\textrm{and} \nonumber\\
  &u_k^{MMSE}& = \frac{R_k h_k^2}{\sigma^2} \left(1-\sum_{c=1}^{C} \alpha^{(c)} \frac{\tilde{\gamma}^{* (c)}}{1+\tilde{\gamma}^{* (c)}} \right)
  \frac{f(\tilde{\gamma}^{* (c)})}{ \tilde{\gamma}^{* (c)}} \nonumber\\ &&\hspace{2cm} {\textrm{for}} \ \ \sum_{c=1}^{C} \alpha^{(c)}
  \frac{\tilde{\gamma}^{* (c)}}{1+\tilde{\gamma}^{* (c)}}< 1 . \label{eq21-3}
\end{eqnarray} }}
Note that, based on the above equations, we have
${{u}_k^{MMSE}>{u}_k^{DE}>{u}_k^{MF}}$. This means that the MMSE
reciever achieves the highest utility as compared to the
decorrelator and the matched filter. Also, the network capacity
(i.e., the number of users that can be admitted into the network) is
the highest when the MMSE detector is used. For the specific case of
no delay constraints, $\tilde{\gamma}^{* (c)}=\gamma^*$ for all $c$.

We can observe from \eqref{eq21-1}--\eqref{eq21-3} that the presence
of users with stringent delay requirements results in a reduction in
the utility of \emph{all} the users in the network. A stringent
delay requirement results in an increase in the user's target SIR
(remember $ \tilde{\gamma}_k^*= \max\{\tilde{\gamma}_k,\gamma^*\}$).
Since $\frac{f(\gamma)}{\gamma}$ is maximal when $\gamma=\gamma^*$,
a target SIR larger than $\gamma^*$ results in a reduction in the
utility of the corresponding user. In addition, because of the
higher target SIR for this user, other users in the network
experience a higher level of interference and hence are forced to
transmit at a higher power which in turn results in a reduction in
their utilities (except for the decorrelator, in which case the
multiple-access interference is completely removed). Also, since $
\tilde{\gamma}_k^*\geq \gamma^*$ and $\sum_{c=1}^{C}
\alpha^{(c)}=\alpha$, the presence of delay-constrained users causes
a reduction in the system capacity (again, except for the
decorrelator). We will demonstrate these losses in
Section~\ref{numerical results} using numerical results.

\section{Delay Model for the Finite Backlog Case}\label{delay model
finite}

So far, we have assumed that there are infinitely many packets for
transmission at each user terminal. Hence, we have focused on the
transmission delay. Now, we extend the analysis to consider the case
in which the packet arrival rate is finite. For this case, we take
into account the queueing delay as well. We specify the QoS
constraints of user $k$ by $(r_k,D_k)$ where $r_k$ is the average
source rate and $D_k$ is the upper bound on average delay. The delay
in this case includes both queuing and transmission delays. The
incoming traffic is assumed to have a Poisson distribution with
parameter $\lambda_k$ which represents the average packet arrival
rate. Since each packet consists of $M$ bits, the source rate $r_k$
(in bit per second) is given by
\begin{equation}\label{eq30}
    r_k= M \lambda_k .
\end{equation}
As before, we assume that the user keeps retransmitting a packet
until the packet is received at the access point without any errors.
The retransmissions are assumed to be independent. The incoming
packets are assumed to be stored in a queue and transmitted in a
first-in-first-out (FIFO) fashion. The packet transmission time for
user $k$ is defined as
\begin{equation}\label{eq31}
    \tau_k = \frac{M}{R_k} \ .
\end{equation}

We can represent the combination of user $k$'s queue and wireless
link as an M/G/1 queue where the traffic is Poisson with parameter
$\lambda_k$ (in packets per second) and the service time, $S_k$, has
the following probability mass function (PMF):
\begin{equation}\label{eq32}
    \textrm{Pr}\{S_k=m\tau_k\}= f(\gamma_k)
    \left(1-f(\gamma_k)\right)^{m-1}  \ \ \ \textrm{for} \ m=1, 2,
    \cdots
\end{equation}
As a result, the service rate, $\mu_k$, is given by
\begin{equation}\label{eq34}
    \mu_k=\frac{1}{\mathbb{E}\{S_k\}}= \frac{f(\gamma_k)}{\tau_k}
    ,
\end{equation}
and the load factor $\rho_k=\frac{\lambda_k}{\mu_k}=\frac{\lambda_k
\tau_k}{f(\gamma_k)}$.

To keep the queue of user $k$ stable, we must have $\rho_k<1$ or
$f(\gamma_k)>\lambda_k \tau_k$. Now, let $W_k$ be a random variable
representing the total packet delay for user $k$. This delay
includes the time the packet spends in the queue as well as the
service time. It can be shown that, for the M/G/1 queue considered
here, the average wait time (including the queuing and service time)
for user $k$ is given by (see \cite{MeshkatiDelayTcomm})
\begin{equation}\label{eq36}
    \bar{W}_k = \tau_k \left(\frac{1-\frac{\lambda_k
    \tau_k}{2}}{f(\gamma_k)-\lambda_k \tau_k}\right) \ \ \
    \textrm{with} \ f(\gamma_k)>\lambda_k \tau_k .
\end{equation}
We require the average delay for user $k$'s packets to be less than
or equal to $D_k$, i.e.,
\begin{equation}\label{eq36b}
    \bar{W}_k \leq D_k .
\end{equation}
Note that $D_k$ cannot be smaller than the transmission time
$\tau_k$. It can be shown again that the delay constraint in
\eqref{eq36b} translates into a lower bound on the output SIR, i.e.,
\begin{equation}\label{eq38b}
 \gamma\geq \hat{\gamma}_k
\end{equation}
 where
\begin{equation}\label{eq39}
    \hat{\gamma}_k=f^{-1}(\hat{\eta}_k)
\end{equation}
with $\hat{\eta}_k=\lambda_k \tau_k + \frac{\tau_k}{D_k}
-\frac{\lambda_k \tau_k^2}{2D_k}$ (again, see
\cite{MeshkatiDelayTcomm}).

\section{Power and Rate Control Game with Delay Constraints}\label{PRCG}

Consider the non-cooperative joint power and rate control game
(PRCG) $\hat{G}=[\mathcal{K}, \{\hat{A}_k\}, \{u_k\}]$ where
$\mathcal{K}=\{1,2,\cdots,K\}$ is the set of users,
$\hat{A}_k=[0,P_{max}]\times[0,B]$ is the strategy set for user $k$
with a strategy corresponding to a choice of transmit power and
transmit rate, and $u_k$ is the utility function for user $k$ given
by \eqref{eq4}. Here, $B$ is the system bandwidth. Each user chooses
its transmit power and rate in order to maximize its own utility
while satisfying its delay QoS requirements. The resulting
delay-constrained power and rate control game can be expressed as
\begin{equation}\label{eq43}
   \max_{p_k, R_k} \ u_k \ \ \ \ \textrm{s.t.} \ \
   \gamma_k \geq \hat{\gamma}_k \ \ \textrm{and} \ \ \ \frac{r_k}{R_k} < \frac{ \frac{D_k R_k}{M}-1
    }{\frac{D_k R_k}{M} -\frac{1}{2}}
\end{equation}
for $k=1, \cdots , K$ where $\hat{\gamma}_k=f^{-1}(\hat{\eta}_k)$
and
\begin{equation}\label{eq43b}
    \hat{\eta}_k=\frac{r_k}{R_k} + \frac{M}{D_k R_k} -\frac{M r_k}{2D_k
    R_k^2} \ .
\end{equation}
The second constraint in \eqref{eq43} is to make sure that
$\hat{\eta}_k<1$. Note that the output SIR $\gamma_k$ depends on
both $p_k$ and $R_k$.

Let us define
\begin{equation} \label{eq50}
\Omega_k^*=\left(\frac{M}{D_k}\right) \frac{1+D_k\lambda_k +\sqrt{1+
D_k^2 \lambda_k^2 +2(1-f^*)D_k \lambda_k}}{2f^*} \ .
\end{equation}
where $f^*=f(\gamma^*)$ with $\gamma^*$ being the (positive)
solution of $f(\gamma)=\gamma f'(\gamma)$. We now state the
following proposition.\vspace{0.1cm}

\begin{proposition}\label{prop2}
If  $\sum_{k=1}^K \frac{1}{1+\frac{B}{\Omega_k^* \gamma^*}} < 1$,
then the proposed delay-constrained power and rate control game has
at least one Nash equilibrium given by $(p_k^*, R_k^*)$,  for
$k=1,\cdots,K$, where $R_k^*=\Omega_k^*$ and $p_k^*$ is the transmit
power that results in an output SIR equal to $\gamma^*$.
Furthermore, when there are more than one Nash equilibrium, $(p_k^*,
\Omega_k^*)$ is the most \emph{efficient} one.
\end{proposition}

\begin{thmproof}{Proof:}
See \cite{MeshkatiDelayTcomm} for the proof.\vspace{0.1cm}
\end{thmproof}

We now define the ``size" of user $k$ as
\begin{equation}\label{eq54}
    \Phi_k^* = \frac{1}{1+\frac{B}{\Omega_k^* \gamma^*}} \ .
\end{equation}
Therefore, the feasibility condition in Proposition~\ref{prop2} can
be written as \vspace{-0.3cm}
\begin{equation}\label{eq55}
    \sum_{k=1}^K \Phi_k^* < 1 .
\end{equation}
The size of a user is basically an indication of the amount of
network resources consumed by that user. Note that the QoS
requirements of user $k$ (i.e., its source rate $r_k$ and delay
constraint $D_k$) uniquely determine $\Omega_k^*$ through
\eqref{eq50} and, in turn, determine the size of the user (i.e.,
$\Phi_k^*$) through \eqref{eq54}. A larger source rate or a tighter
delay constraint for a user increases the size of the user. The
network can accommodate a set of users if and only if their total
size is less than 1. In Section~\ref{numerical results}, we use this
framework to study the tradeoffs among throughput, delay, network
capacity and energy efficiency.

\section{Numerical Results}\label{numerical results}

Let us first consider the infinite backlog case as discussed in
Sections~\ref{delay model infinite}--\ref{multiclass}. Let us
consider a DS-CDMA system with processing gain 100. We assume that
each packet contains 100 bits (i.e., $M=100$). The transmission
rate, $R$, is $100$kbps. A useful example for the efficiency
function is ${f(\gamma)= (1- e^{-\gamma})^M}$. This serves as an
approximation to the packet success rate that is very reasonable for
moderate to large values of $M$. We use this efficiency function for
our simulations. Using this, with $M=100$, we have $\gamma^*=6.48 =
8.1$dB.

\begin{figure}
\begin{center}
\leavevmode \hbox{\epsfysize=5.3cm \epsfxsize=8cm
\epsffile{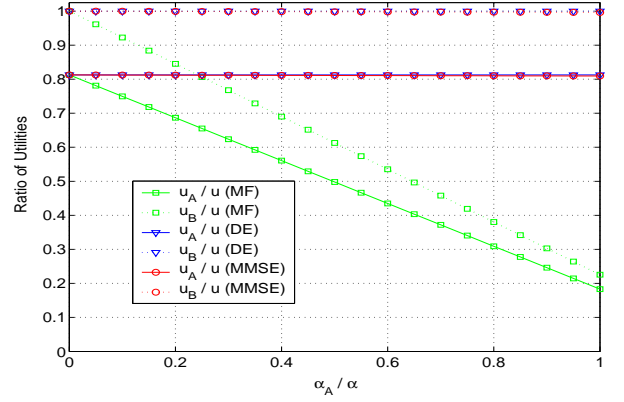}}\vspace{-0.3cm}
\end{center}
\caption{Loss in utility due to presence of users with stringent
delay requirements ($\alpha=0.1$).} \label{fig2}\vspace{-0.2cm}
\end{figure}

We consider a network where the users can be divided into two
classes: delay sensitive (class $A$) and delay tolerant (class $B$).
For users in class $A$, we choose $L_A=1$ and $\beta_A=0.99$ (i.e.,
delay sensitive). For users in class $B$, we let $L_B=3$ and
$\beta_B=0.90$ (i.e., delay tolerant). Based on these choices,
$\tilde{\gamma}^*_A=9.6$dB and $\tilde{\gamma}^*_B=\gamma^*=8.1$dB.
Without loss of generality and to keep the comparison fair, we also
assume that all the users are the same distance from the access
point. The system load is $\alpha$ (i.e., $\frac{K}{N}=\alpha$) and
we let $\alpha_A$ and $\alpha_B$ represent the load corresponding to
class $A$ and class $B$ users, respectively, with ${\alpha_A +
\alpha_B =\alpha}$. We first consider a lightly loaded network with
$\alpha=0.1$ (see Fig. \ref{fig2}). To demonstrate the performance
loss due to the presence of users with stringent delay requirements
(i.e., class $A$), we plot ${u_A}/{u}$ and ${u_B}/{u}$ as a function
of the fraction of the load corresponding to class $A$ users (i.e.,
${\alpha_A}/{\alpha}$). Here, $u_A$ and $u_B$ are the utilities of
users in class $A$ and class $B$, respectively, and $u$ represents
the utility of the users if they all had loose delay requirements.
Fig. \ref{fig2} shows the loss in utility for the matched filter,
the decorrelator, and the MMSE detector. We observe from the figure
that for the matched filter both classes of users suffer
significantly due to the presence of delay sensitive traffic. For
example, when half of the users are delay-sensitive, the utilities
achieved by class $A$ and class $B$ users are, respectively, 50\%
and 60\% of the utilities for the case of no delay constraints. For
the decorrelator, only class $A$ users suffer and the reduction in
utility is smaller than that of the matched filter. For the MMSE
detector, the reduction in utility for class $A$ users is similar to
that of the decorrelator, and the reduction in utility for class $B$
is negligible.

\begin{figure}
\begin{center}
\leavevmode \hbox{\epsfysize=5.3cm \epsfxsize=8cm
\epsffile{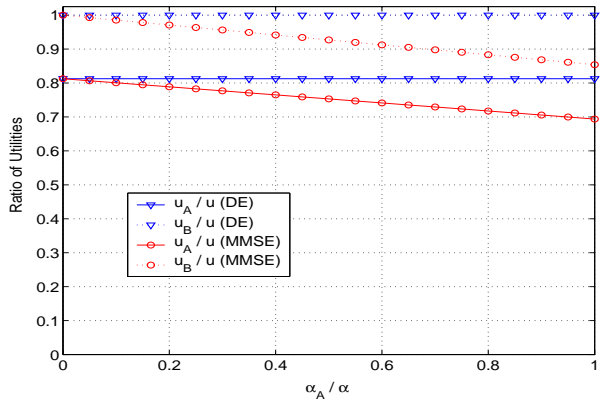}}\vspace{-0.3cm}
\end{center}
\caption{Loss in utility due to presence of users with stringent
delay requirements ($\alpha=0.9$).} \label{fig3}\vspace{-0.2cm}
\end{figure}

We repeat the experiment for a highly loaded network with
$\alpha=0.9$ (see Fig. \ref{fig3}). Since the matched filter cannot
handle such a significant load, we have shown the plots for the
decorrelator and MMSE detector only. We observe from Fig.~\ref{fig3}
that because of the higher system load, the reduction in the
utilities is more significant for the MMSE detector compared to the
case of $\alpha=0.1$. It should be noted that for the decorrelator
the reduction in utility of class $A$ users is independent of the
system load. This is because the decorrelator completely removes the
multiple-access interference.

\begin{figure}
\begin{center}
\leavevmode \hbox{\epsfysize=5.3cm \epsfxsize=8cm
\epsffile{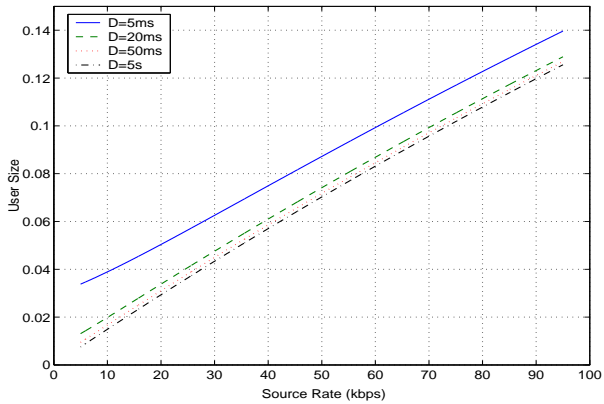}}\vspace{-0.3cm}
\end{center}
\caption{User size, $\Phi^*$, as a function of source rate for
different delay requirements.} \label{sizevsrate}\vspace{-0.2cm}
\end{figure}

\begin{figure}
\begin{center}
 \leavevmode \hbox{\epsfysize=5.3cm \epsfxsize=8cm
\epsffile{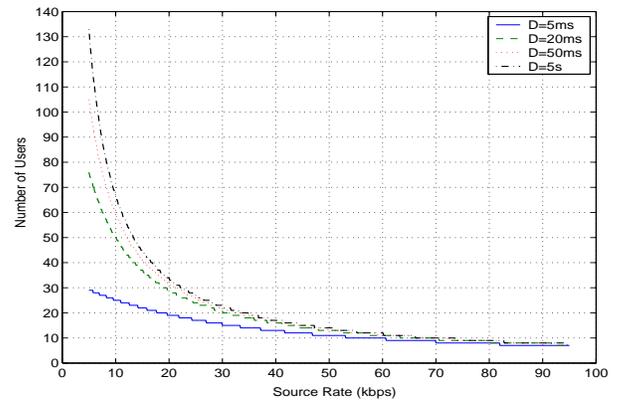}}\vspace{-0.3cm}
\end{center} \caption{Network capacity as a function of source rate
for different delay requirements.}
\label{numusersvsrate}\vspace{-0.2cm}
\end{figure}

\begin{figure}
\centering \leavevmode \hbox{\epsfysize=5.3cm \epsfxsize=8cm
\epsffile{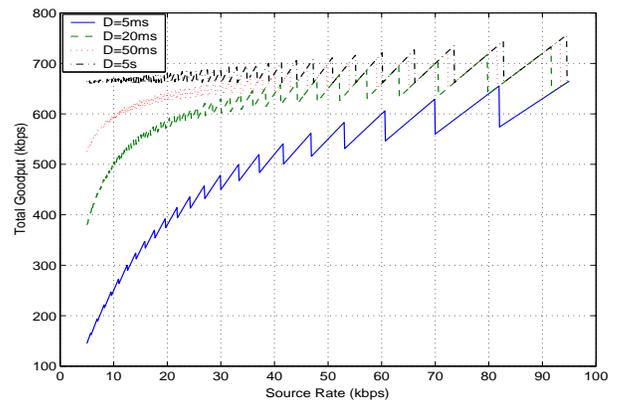}}\vspace{-0.1cm} \caption{Total
goodput as a function of source rate for different delay
requirements.} \label{totalgpvsrate}\vspace{-0.2cm}
\end{figure}

We now present simulation results for the finite backlog case as
discussed in Sections~\ref{delay model finite} and \ref{PRCG}. Let
us consider the uplink of a DS-CDMA system with a total bandwidth of
5MHz (i.e. $B=5$MHz). As explained in Section~\ref{PRCG}, the QoS
parameters of a user define a ``size" for that user, denoted by
$\Phi_k^*$ given by \eqref{eq54}. Before a user starts transmitting,
it must announce its size to the access point. Based on the
particular admission policy, the access point decides whether or not
to admit the user. Throughout this section, we assume that the
admitted users choose the transmit powers and rates that correspond
to their efficient Nash equilibrium (see Proposition~\ref{prop2}).
Fig.~\ref{sizevsrate} shows the size of a user as a function of the
user's source rate and for different delay requirements. It is seen
that the higher the source rate and the tighter the delay
requirement, the larger the size. Now, let us assume that all users
in the network have the same QoS requirements, which means that all
the users have the same size. Based on \eqref{eq55}, we can
calculate the maximum number of users whose QoS requirements can be
accommodated (i.e., network capacity). Fig.~\ref{numusersvsrate}
shows the network capacity as a function of the source rate for
different delay requirements. As the source rate increases and the
delay bound becomes tighter, the number of users that can be
accommodate by the network reduces. Eventually, as the source rate
becomes very large, only one user can be accommodated by the
network. We can also plot the total goodput (i.e., reliable
throughput) of the network. Fig.~\ref{totalgpvsrate} shows the total
goodput as a function of the source rate for different delay
requirements.

\section{Conclusions} \label{conclusion}
We have studied the energy-delay tradeoffs using a game-theoretic
framework. A non-cooperative game is proposed in which each user
chooses its transmit power, and possibly its transmission rate, to
maximize its own utility (in bits per Joule) while satisfying its
delay QoS requirements. The Nash equilibrium solution for the
proposed game is presented. We have shown that the presence of
delay-sensitive users results in significant losses in the network
utility and capacity, and have quantified the losses. The tradeoffs
among throughput, delay, network capacity and energy efficiency have
also been analyzed.
\section*{Acknowledgment}

This research was supported in part by the National Science
Foundation under Grant ANI-03-38807.


\begin{thebibliography}{10}
{\scriptsize{
\bibitem{Collins99}
B.~Collins and R.~Cruz, ``Transmission policies for time varying
channels with
  average delay constraints,'' {\em Proceedings of the $37^{th}$ Annual
  Allerton Conference on Communication, Control, and Computing}, Monticello,
  IL, October 1999.

\bibitem{Prabhakar01}
B.~Prabhakar, E.~Uysal-Biyikoglu, and A.~El~Gamal,
``Energy-efficient
  transmission over a wireless link via lazy packet scheduling,'' {\em
  Proceedings of $20^{th}$ Annual Joint Conference of the IEEE Computer and
  Communications Societies (INFOCOM)}, Anchorage, AK, April 2001.

\bibitem{Berry02}
R.~A. Berry and R.~G. Gallager, ``Communication over fading channels
with delay
  constraints,'' {\em IEEE Transactions on Information Theory}, vol.~48,
  pp.~1135--1149, May 2002.

\bibitem{Uysal02}
E.~Uysal-Biyikoglu and A.~El~Gamal, ``Energy-efficient packet
transmission over
  multiaccess channel,'' {\em Proceedings of IEEE International Symposium on
  Information Theory (ISIT)}, Lausanne, Switzerland, June/July 2002.

\bibitem{Fu03}
A.~Fu, E.~Modiano, and J.~Tsitsiklis, ``Optimal energy allocation
for
  delay-constrained data transmission over a time-varying channel,'' {\em
  Proceedings of $22^{nd}$ Annual Joint Conference of the IEEE Computer and
  Communications Societies (INFOCOM)}, San Francisco, CA, March/April 2003.

\bibitem{Coleman04}
T.~P. Coleman and M.~M\'{e}dard, ``A distributed scheme for
achieving
  energy-delay tradeoffs with multiple service classes over a dynamically
  varying network,'' {\em IEEE Journal on Selected Areas in Communications
  (JSAC)}, vol.~22, pp.~929--941, June 2004.

\bibitem{Meshkati_TCOMM}
F.~Meshkati, H.~V. Poor, S.~C. Schwartz, and N.~B. Mandayam, ``An
  energy-efficient appraoch to power control and receiver design in wireless
  data networks,'' {\em IEEE Transactions on Communications}, vol.~52,
  pp.~1885--1894, November 2005.

\bibitem{MeshkatiISIT}
F.~Meshkati, H.~V. Poor, and S.~C. Schwartz, ``A non-cooperative
power control
  game in delay-constrained multiple-access networks,'' {\em Proceedings of the
  IEEE International Symposium on Information Theory (ISIT)}, Adelaide,
  Australia, September 2005.

\bibitem{MeshkatiDelayTcomm}
F.~Meshkati, H.~V. Poor, S.~C. Schwartz, and R.~Balan,
``Energy-efficient
  resource {allocation} in wireless networks with quality-of-service
  constraints,'' preprint, Princeton University, 2005.
}}
\end{thebibliography}
\end{document}